\documentclass[namedreferences]{solarphysics}

\usepackage[hyperref,optionalrh]{spr-sola-addons} 
\usepackage{graphicx}        
\usepackage{color}           
\usepackage{breakurl}        

\newcommand{\aap}{    {\it Astron. Astrophys.}}

\newcommand{\apj}{    {\it Astrophys. J.}}
\newcommand{\apjl}{   {\it Astrophys. J. Lett.}}

\newcommand{\solphys}{{\it Solar Phys.}}

\begin{document}

\begin{article}

\begin{opening}

\title{High-resolution Observations of a Flux Rope with the \emph{Interface Region Imaging Spectrograph}}

\author{Ting~\surname{Li}$^{1}$\sep
        Jun~\surname{Zhang}$^{1}$\sep
       }
\runningauthor{Li \& Zhang} \runningtitle{Flux Rope Observed with
IRIS}

   \institute{$^{1}$ Key Laboratory of Solar Activity, National
Astronomical Observatories, Chinese Academy of Sciences, Beijing
100012, China;
                     email: \url{liting@nao.cas.cn}\\
             }

\begin{abstract}
We report the observations of a flux rope at transition region
temperatures with the \emph{Interface Region Imaging Spectrograph}
(IRIS) on 30 August 2014. Initially, magnetic flux cancellation
constantly took place and a filament was activated. Then the bright
material from the filament moved southward and tracked out several
fine structures. These fine structures were twisted and tangled with
each other, and appeared as a small flux rope at 1330 {\AA}, with a
total twist of about 4$\pi$. Afterwards, the flux rope underwent a
counter-clockwise (viewed top-down) unwinding motion around its
axis. Spectral observations of C {\sc ii} 1335.71 {\AA} at the
southern leg of the flux rope showed that Doppler redshifts of
6$-$24 km s$^{-1}$ appeared at the western side of the axis, which
is consistent with the counter-clockwise rotation motion. We suggest
that the magnetic flux cancellation initiates reconnection and some
activation of the flux rope. The stored twist and magnetic helicity
of the flux rope are transported into the upper atmosphere by the
unwinding motion in the late stage. The small-scale flux rope (width
of 8.3$^{\prime\prime}$) had a cylindrical shape with helical field
lines, similar to the morphology of the large-scale CME core (width
of 1.54 $R_{\odot}$) on 2 June 1998. This similarity shows the
presence of flux ropes of different scales on the Sun.

\end{abstract}
\end{opening}

\section{Introduction}
     \label{}

Eruptive solar filaments and coronal mass ejections (CMEs) often
exhibit a clear helical geometry (about 40\% for CMEs; Vourlidas
\emph{et al.}, 2012), indicating the eruption of a magnetic flux
rope. A magnetic flux rope becomes kink-unstable if the twist
exceeds a critical value of 2.5$\pi$$-$3.5$\pi$ (Hood and Priest,
1981; Fan, 2005; T{\"o}r{\"o}k and Kliem, 2005). Then the axis of
the flux rope undergoes writhing motions, and the twist is
transformed into the writhe of the axis, since the magnetic helicity
is essentially conserved (Ji \emph{et al.}, 2003; Rust and LaBonte,
2005). The kink instability can initiate the rise motion of the flux
rope and is considered as one type of mechanism that triggers the
CME and associated activities (T{\"o}r{\"o}k and Kliem, 2003; Liu
\emph{et al.}, 2007; Yan \emph{et al.}, 2014).

However, a basic question about the formation of the flux rope is
not well understood. Some theoretical models suggest that the flux
rope is formed through magnetic reconnection, which converts sheared
arcades into a helical structure (Pneuman, 1983; van Ballegooijen
and Martens, 1989). Joshi \emph{et al.} (2014a) showed an example of
compound flux rope formation via merging of two different flux
ropes, consistent with the tether cutting magnetic reconnection
scenario (Moore \emph{et al.}, 2001). Cheng \emph{et al.} (2014)
suggested that magnetic reconnection was responsible for the
formation of a double-decker magnetic flux rope. Another possibility
is the emergence of the flux rope from below the photosphere into
the corona. The helical structure is formed deep in the convection
zone before its emergence (Rust and Kumar, 1994; Lites, 2005).
Okamoto \emph{et al.} (2008) analyzed the vector magnetic fields on
the photosphere under a filament and concluded that a helical flux
rope was emerging from below the photosphere.

Based on the nonlinear force-free field (NLFFF) extrapolation
technique, Guo \emph{et al.} (2010) and Canou and Amari (2010)
showed the presence of twisted flux ropes and found that the
location of the magnetic dips within the flux ropes agrees with the
observed filament in H$\alpha$ images. Moreover, the formation and
eruption of flux ropes have been simulated by many authors (Fan and
Gibson, 2004; Amari \emph{et al.}, 2010). Aulanier \emph{et al.}
(2010) carried out the magnetohydrodynamic (MHD) simulation and
found that the photospheric flux-cancellations in a bald-patch
separatrix and tether-cutting coronal reconnection are key
mechanisms for the formation of flux ropes.

Recently, the direct observations of flux ropes in the low corona
have been reported (Li and Zhang, 2013a, 2013c, 2014; Chen \emph{et
al.}, 2014a, 2014b; Song \emph{et al.}, 2014; Joshi \emph{et al.},
2014b) by using the extreme ultraviolet (EUV) data from the
\emph{Atmospheric Imaging Assembly} (AIA; Lemen \emph{et al.}, 2012)
onboard the \emph{Solar Dynamics Observatory} (SDO; Pesnell \emph{et
al.}, 2012). Kumar \emph{et al.} (2010) presented a successive
activation of magnetic flux ropes in chromospheric and EUV lines and
concluded that such activation plays an important role in triggering
the flare. Srivastava \emph{et al.} (2010) reported the observations
of a flux rope with a high twist angle of 12$\pi$ in the channels of
Ca {\sc ii} H (3968 {\AA}) and 171 {\AA}. The recently launched
\emph{Interface Region Imaging Spectrograph} (IRIS; De Pontieu
\emph{et al.}, 2014) mission is now providing observations of the
transition region (TR) and chromosphere with remarkable spatial and
spectral resolution. The TR is the interface between the
chromosphere and the corona, where the temperature rapidly rises
from 25000 K to 1 MK. In this work, we present the observations of a
flux rope at TR temperatures based on IRIS and SDO data.

\section{Observations and Data Analysis} 
      \label{}

\begin{figure}
   \centerline{\includegraphics[width=1.0\textwidth,clip=]{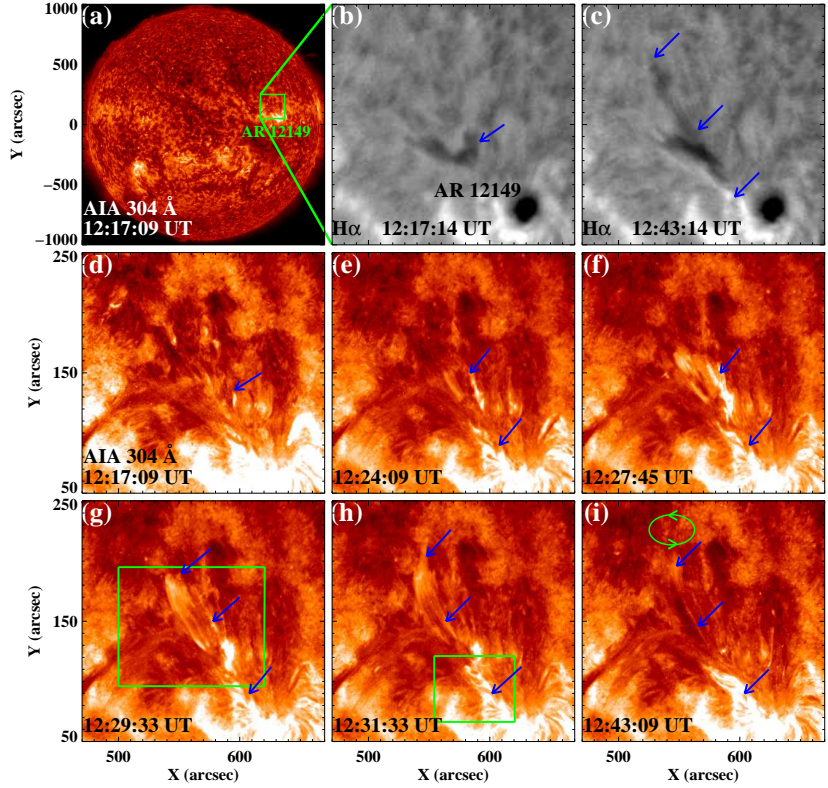}
              }
   \caption{SDO$/$AIA 304 {\AA} and GONG H$\alpha$ images showing the evolution of the erupting
filament on 30 August 2014 (a 304 {\AA} movie is also available as
electronic supplementary material). Blue arrows point to the
analyzed filament. The green rectangle in panel (g) denotes the
field of view (FOV) of Figures 3(a)-(b), and the green rectangle in
panel (h) displays the FOV of the images in Figures 4 and 5. The
green ellipse in panel (i) shows the counterclockwise rotation of
the filament.}
              \label{Fig1}%
    \end{figure}

IRIS records spectra over three wavelength bands of the far
ultraviolet (1332$-$1358 {\AA} and 1389$-$1406 {\AA}) and the near
ultraviolet (2782$-$2834 {\AA}), and also obtains slit-jaw images
(SJIs) centered at 1330, 1400, 2796 and 2832 {\AA}, with
0.33$^{\prime\prime}$$-$0.4$^{\prime\prime}$ spatial resolution (De
Pontieu \emph{et al.}, 2014). For the analyzed event, the IRIS
observation was taken from 11:12 UT to 17:13 UT on 30 August 2014.
The SJIs are taken in the 1330 and 2796 {\AA} channels. The 1330
{\AA} channel best shows the flux rope and we focus on this channel
in this study. The 1330 {\AA} filter records emission coming from
the C {\sc ii} 1335.71 {\AA} line and the UV continuum. The C {\sc
ii} 1335.71 {\AA} line is formed in the lower TR and corresponds to
a temperature of about 10$^{4.4}$ K (Tian \emph{et al.}, 2014; Li
\emph{et al.}, 2014). The SJIs at 1330 {\AA} have a cadence of
$\approx$ 19 s and a sampling of 0.166$^{\prime\prime}$
pixel$^{-1}$. The spectral observations covering the strong emission
line of C {\sc ii} 1335.71 {\AA} are analyzed in detail. The
spectral data are taken in a sit-and-stare mode with 8 s exposure
time and 9 s cadence. The present study uses IRIS level 2 data
provided by the IRIS team. Dark current removal, flat-field and
geometric correction have been taken into account in the level 2
data. The emission in the the IRIS spectral line of C {\sc ii} is
shifted along the IRIS slit, and this offset is achieved by checking
the position of horizontal fiducial lines.

The SDO observations are also used here in order to analyze the
filament evolution. The AIA takes full-disk images in 10 (E)UV
channels at 1.5$^{\prime\prime}$ resolution and high cadence of 12
s. Among these channels, the high-cadence 171 and 304 {\AA}
observations are chosen. The 171 {\AA} channel corresponds to a
temperature of about 0.6 MK (Fe {\sc ix}) and the channel of 304
{\AA} (He {\sc ii}) is at 0.05 MK (O'Dwyer \emph{et al.}, 2010).
Moreover, we also use the full-disk line-of-sight (LOS) magnetic
field data from the \emph{Helioseismic and Magnetic Imager} (HMI;
Scherrer \emph{et al.}, 2012) onboard SDO, with a cadence of
$\approx$ 45 s and a sampling of 0.5$^{\prime\prime}$ pixel$^{-1}$.
NSO$-$GONG H$\alpha$ data are used to investigate the chromospheric
configuration of the filament. GONG collects H$\alpha$ data at seven
sites around the world with a spatial resolution of
1.0$^{\prime\prime}$ pixel$^{-1}$ and a cadence of around 1 min
(Harvey \emph{et al.}, 2011). The morphology of the flux rope
reminds us of the CME event occurring on 2 June 1998. The
\emph{Large Angle and Spectrometric Coronagraph} (LASCO; Brueckner
\emph{et al.}, 1995) experiment on board the \emph{Solar and
Heliospheric Observatory} (SOHO) records the CME with a ``net-like
cylinder" shape, similar to the flux rope observed by the IRIS.

\section{Results}
\subsection{Failed Eruption of the Filament}


The filament analyzed here is located in the north of NOAA active
region (AR) 12149 near the coordinates (600$^{\prime\prime}$,
50$^{\prime\prime}$) around 12:20 UT on 30 August 2014 (Figure
1(a)). It has a length of about 50$^{\prime\prime}$, the typical
length of a mini-filament (Zheng \emph{et al.}, 2012; Kumar and Cho,
2014). Before the activation, the filament appeared faint both in
H$\alpha$ and EUV wavelengths (Figures 1(b) and (d)). At about 12:20
UT, the filament was activated and then it could be clearly observed
at 304 {\AA}. The associated EUV brightenings appeared at the west
of the filament and increased until 12:27 UT (Figures 1(e)-(f);
movies in 304 {\AA} and 171 {\AA} are available as electronic
supplementary materials). The eruptive material moved northward with
a velocity of about 100 km s$^{-1}$, and the northern part of the
filament showed an evident counterclockwise rotation around the
filament axis (Figures 1(g)-(i)). The filament was not clearly
observed in H$\alpha$ images during the eruption process. This might
be caused by the heating of the filament material. Until 12:43 UT,
the filament material was cooled and thus the filament appeared
again in H$\alpha$ images (Figure 1(c)).

\begin{figure}
   \centerline{\includegraphics[width=1.0\textwidth,clip=]{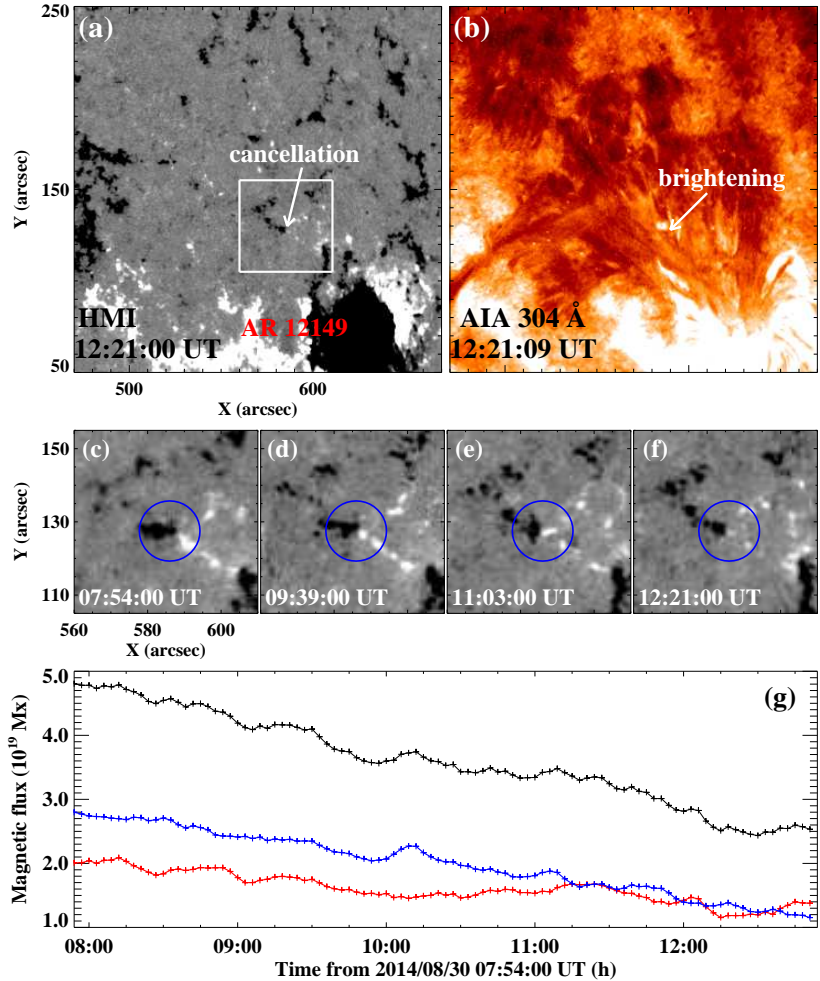}
              }
   \caption{SDO$/$AIA 304 {\AA} image, HMI LOS magnetograms and temporal variations of the magnetic
flux showing the brightening and cancellation process of magnetic
fields nearby the filament. The white square in panel (a) displays
the FOV of panels (c)-(f). Blue circles outline the area where the
temporal evolution of magnetic flux is calculated. The black profile
in panel (g) denotes the evolution of total magnetic flux and the
red and blue profiles respectively denote the positive and negative
magnetic flux.}
              \label{Fig2}%
    \end{figure}

Starting from about 08:00 UT, magnetic flux cancellation constantly
took place underlying the northern part of the filament (Figure 2).
The temporal variations of the cancelling magnetic flux in HMI
magnetograms were measured and are displayed in Figure 2(g). We
derotated all the magnetograms differentially to a reference time
(11:10 UT). The area in which we calculated the magnetic flux is the
blue circle in Figures 2(c)-(f). After selection, the pixels with
magnetic fields weaker than 10.2 gauss (G) (the noise level
determined by Liu \emph{et al.}, 2012) are eliminated. The temporal
profiles of positive and negative magnetic flux show a consistent
decreasing trend. The total magnetic flux decreased from
4.8$\times$10$^{19}$ maxwell (Mx) at 07:54 UT to
2.5$\times$10$^{19}$ Mx at 12:55 UT. The average rate of flux
cancellation was approximately 1.3$\times$10$^{15}$ Mx s$^{-1}$. At
12:21 UT, the initial EUV brightening was observed at the location
of the cancelled flux and the filament started to erupt (Figure
2(b)).

\begin{figure}
   \centerline{\includegraphics[width=1.0\textwidth,clip=]{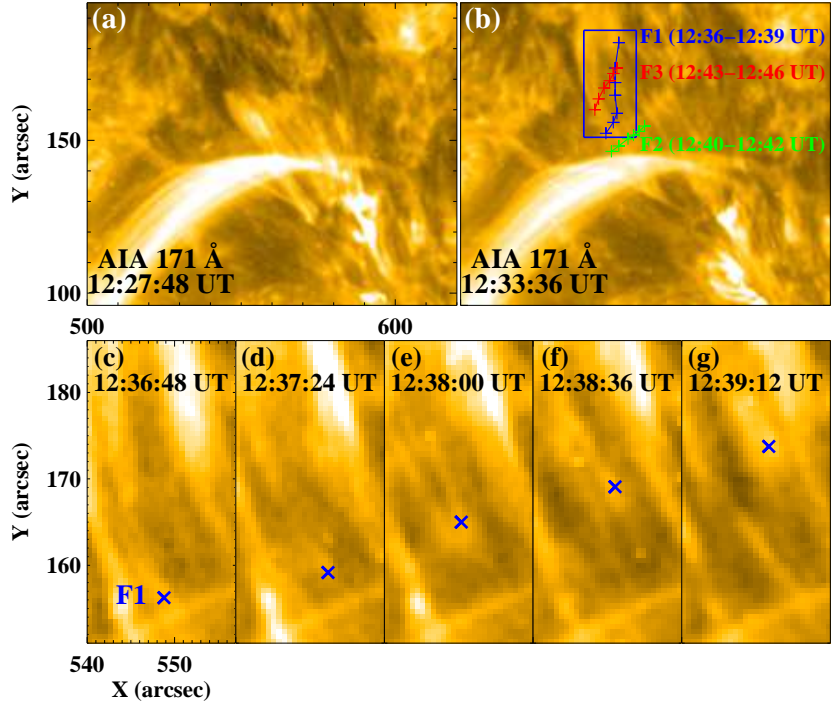}
              }
   \caption{SDO$/$AIA 171 {\AA} images showing the counterclockwise
rotation of the filament (a movie in 171 {\AA} is available as
electronic supplementary material). The plus symbols in panel (b)
mark the tracks of rotating features (``F1"$-$``F3"). The times in
parentheses are the start and end tracking times of the
corresponding features. The blue rectangle in panel (b) denotes the
FOV of panels (c)-(g). The cross symbols in panels (c)-(g) indicate
the positions of ``F1" at different times.}
              \label{Fig3}%
    \end{figure}

To reveal the rotation of the northern filament more clearly, we
tracked three moving features (``F1"$-$``F3" in Figure 3(b)) between
12:36 UT to 12:46 UT. As an example, the evolution of ``F1" was
shown in Figures 3(c)-(g), exhibiting a helical upward motion. Each
track of moving features from the eastern edge to the west
corresponds to a rotation angle of about $\pi$, and thus the total
twist angle was approximately 3$\pi$ in about 10 min. The average
angular speed of the rotating plasma was about 15.7$\times$10$^{-3}$
rad s$^{-1}$. The filament material ultimately fell back to its
initial location from 12:50 UT, and the eruption was failed.

\begin{figure}
   \centerline{\includegraphics[width=1.0\textwidth,clip=]{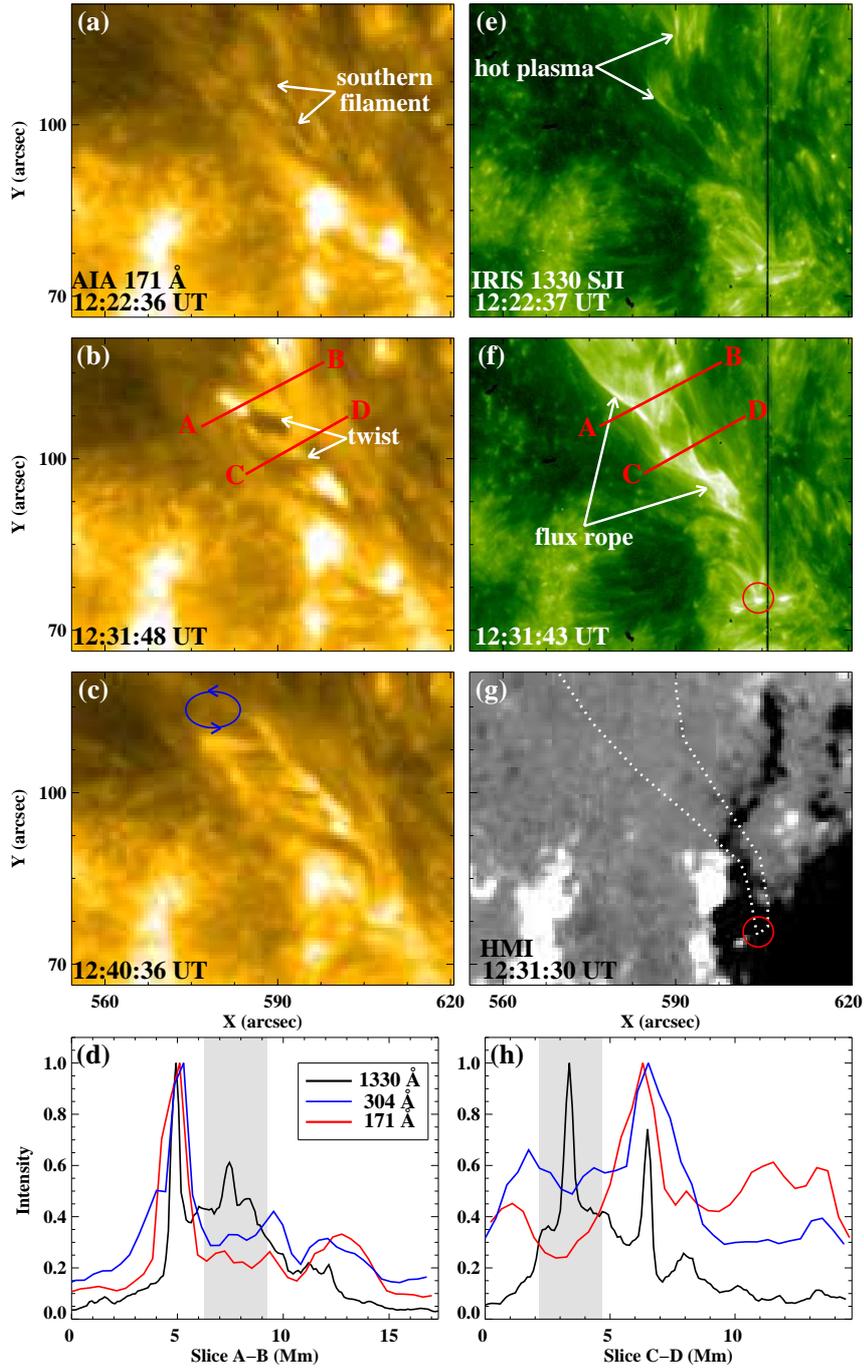}
              }
   \caption{Panels (a)-(c): helical evolution
of the southern filament. The blue ellipse in panel (c) shows the
counterclockwise rotation of the filament. Panels (e)-(g):
appearance of the flux rope and the corresponding HMI magnetogram.
The red circles in panels (f)-(g) outline the southern end of the
flux rope. White dotted contour in panel (g) outlines the location
of the flux rope. Panels (d) and (h): normalized intensity of the
multi-wavelength cuts along slices ``A$-$B" and ``C$-$D" (red
straight lines in panels (b) and (f)), respectively.}
              \label{Fig4}%
    \end{figure}

The southern part of the filament showed an obvious helical
evolution during the eruption process (Figure 4). Before the
appearance of EUV brightening, the southern filament looked linear
and showed no helical shape (Figure 4(a)). Afterwards, it was
activated and the helical structure gradually appeared (panel (b)).
After 12:35 UT, the southern filament showed a rotation motion in
the counterclockwise direction (panel (c)). IRIS observations showed
that a flux rope with a helical shape appeared at the site of the
southern filament (panels (e)-(f)). The comparison of the HMI
magnetogram with the 1330 {\AA} image shows that the southern end of
the flux rope was rooted in the negative polarity fields of the AR
(panels (f)-(g)). In order to compare the multi-wavelength
appearances of the flux rope and filament material, the intensity
plots along the cuts ``A$-$B" and ``C$-$D" (red straight lines in
panels (b) and (f)) at 12:31:48 UT are presented in the bottom
panels of Figure 4. All the cuts along slice ``A$-$B" showed a
strongest emission at about 5 Mm (first peak in panel (d)). The
emission of 1330 {\AA} had a secondary peak at about 7.5 Mm (gray
section in panel (d)). However, the cuts of 304 {\AA} and 171 {\AA}
show much weaker emissions at about 7.5 Mm. In the 1330 {\AA} cut
along slice ``C$-$D", the section corresponding to the filament
material, \emph{i.e.}, the gray section in panel (h), showed an
intensity enhancement while in the 304 {\AA} and 171 {\AA} cuts,
there was a significantly lower emission at the corresponding
section. This indicates a striking anti-correlation between the 1330
{\AA} and EUV line intensities at the location of the filament,
\emph{i.e.}, peaks of the former correspond to minimum of the
latter.

\subsection{Flux Rope Observed by the IRIS}
\begin{figure}
   \centerline{\includegraphics[width=1.0\textwidth,clip=]{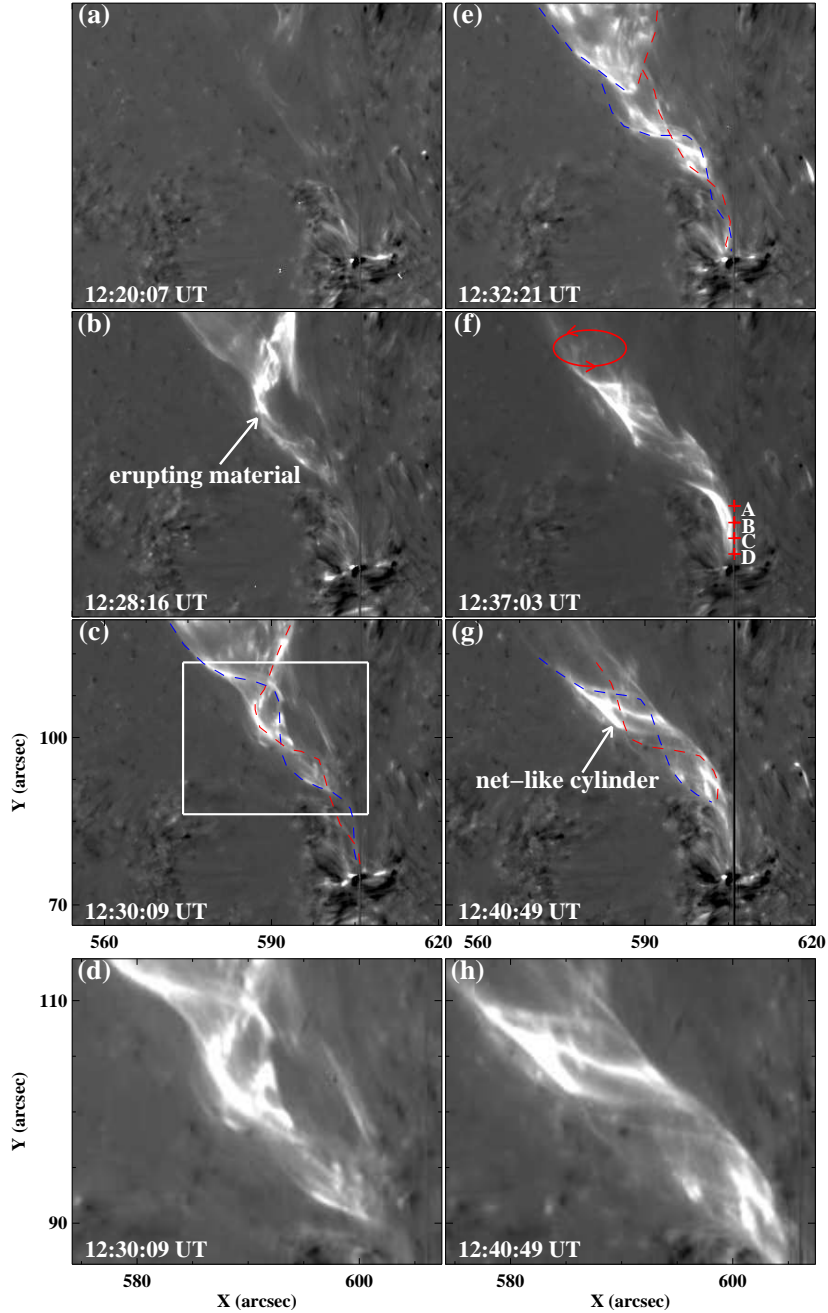}
              }
   \caption{IRIS
base difference 1330 {\AA} slit-jaw images showing the appearance of
a flux rope and its helical evolution (a movie is available as
electronic supplementary material). The base image is at 12:00 UT.
The red and blue dashed curves in panels (c), (e) and (g) outline
the fine-scale structures of the flux rope. The white rectangle in
panel (c) displays the FOV of the zoomed images in panels (d) and
(h). The red ellipse in panel (f) shows the unwinding motion of the
flux rope. The four plus symbols (labeled ``A"$-$``D") along the
slit in panel (f) mark the locations where we perform detailed
analysis of the line profiles in Figure 6.}
              \label{Fig5}%
    \end{figure}

Starting from about 12:20 UT, the erupting material of the filament
gradually moved southwardly and tracked out bright fine-scale
features (Figures 5(a)-(b); a movie is available as electronic
supplementary materials). At 12:28 UT, the initial flux rope looked
like a funnel and the kink could not be clearly observed (Figure
5(b)). Subsequently, the fine threads were tangled and intertwined
with each other. Two individual threads were selected to estimate
the twist of the flux rope and they showed crossing several times in
an image (red and blue dashed curves in Figure 5(c)). We estimated
the amount of twist was about 4$\pi$ according to the geometry of
the threads (panels (c)-(d)). At 12:32 UT, the bright plasma arrived
at the southern end of the flux rope and illuminated the whole flux
rope (panel (e)). The same amount of twist of about 4$\pi$ was
visible at this time from the southern footpoint up to the northern
part. The length of the visible flux rope was approximately
40$^{\prime\prime}$. The flux rope was not stable and underwent a
counterclockwise rotation around its main axis after 12:35 UT
(Figure 5(f)). Initially intertwined strands began to untangle and
the morphology of the flux rope was changed evidently. The rotation
of the flux rope was obviously the unwinding motion of the twisted
magnetic field lines, consistent with the rotation direction of the
filament (Figures 3 and 4). Associated with the unwinding motion,
the flux rope finally evolved into a ``net-like cylinder" (panel
(g)), with the southern leg slightly thinner than top parts. The two
zoomed images at 12:30 UT and 12:41 UT (panels (d) and (h)) clearly
showed the twisted threads of the flux rope. Later on, the intensity
of the flux rope decreased and the flux rope became invisible after
12:48 UT.

\begin{figure}
   \centerline{\includegraphics[width=1.0\textwidth,clip=]{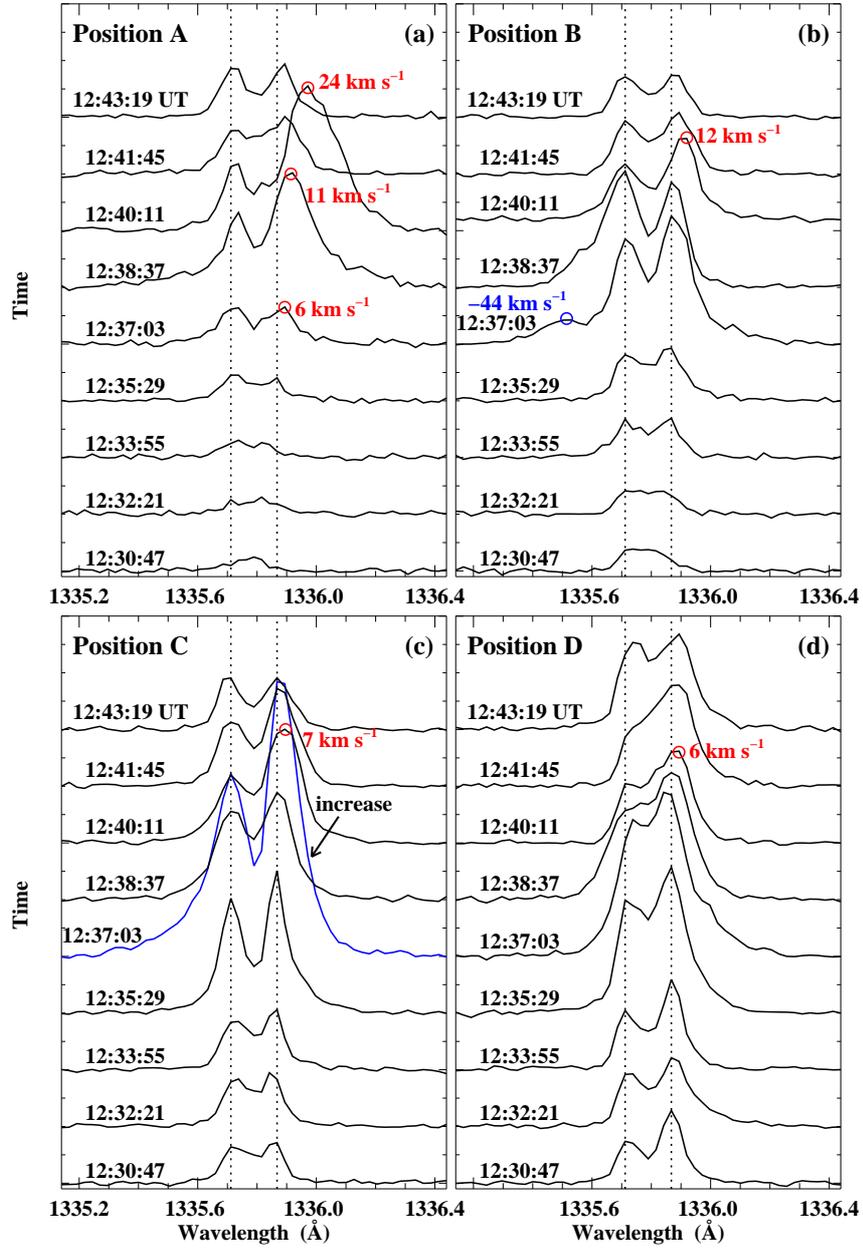}
              }
   \caption{IRIS
line profile evolution of C {\sc ii} 1335.71 {\AA} at the locations
of ``A"$-$``D" marked in Figure 5(f). These profiles cover the
wavelength range from 1335.14 {\AA} to 1336.44 {\AA}. Two dotted
lines in each panel denote two line centers of 1335.71 {\AA} and
1335.86 {\AA}.}
              \label{Fig6}%
    \end{figure}

Figure 6 shows the line profile (C {\sc ii} 1335.71 {\AA}) evolution
at different points (locations ``A"$-$``D" in Figure 5(f)) along the
slit. The slit was located at the southern leg of the flux rope in
the SJIs (Figure 5). In the wavelength range of 1335.14$-$1336.44
{\AA}, the line profiles mainly show two peaks at 1335.71 {\AA} and
1335.86 {\AA}. The emissions at locations ``A" and ``B" between
12:30:47 UT and 12:35:29 UT were relatively weak, compared to those
at locations ``C" and ``D". By checking the SJIs, we suggest that
the weak line profiles of ``A" and ``B" in the early stage mostly
came from the background plasma, and the emissions of southern
locations ``C" and ``D" were from the plasma of the flux rope.
Associated with the gradual expansion of the flux rope, the emission
intensity of northern locations ``A" and ``B" increased between
12:37:03 UT and 12:40:11 UT due to the appearance of the flux rope
at these locations.

\begin{figure}
   \centerline{\includegraphics[width=1.0\textwidth,clip=]{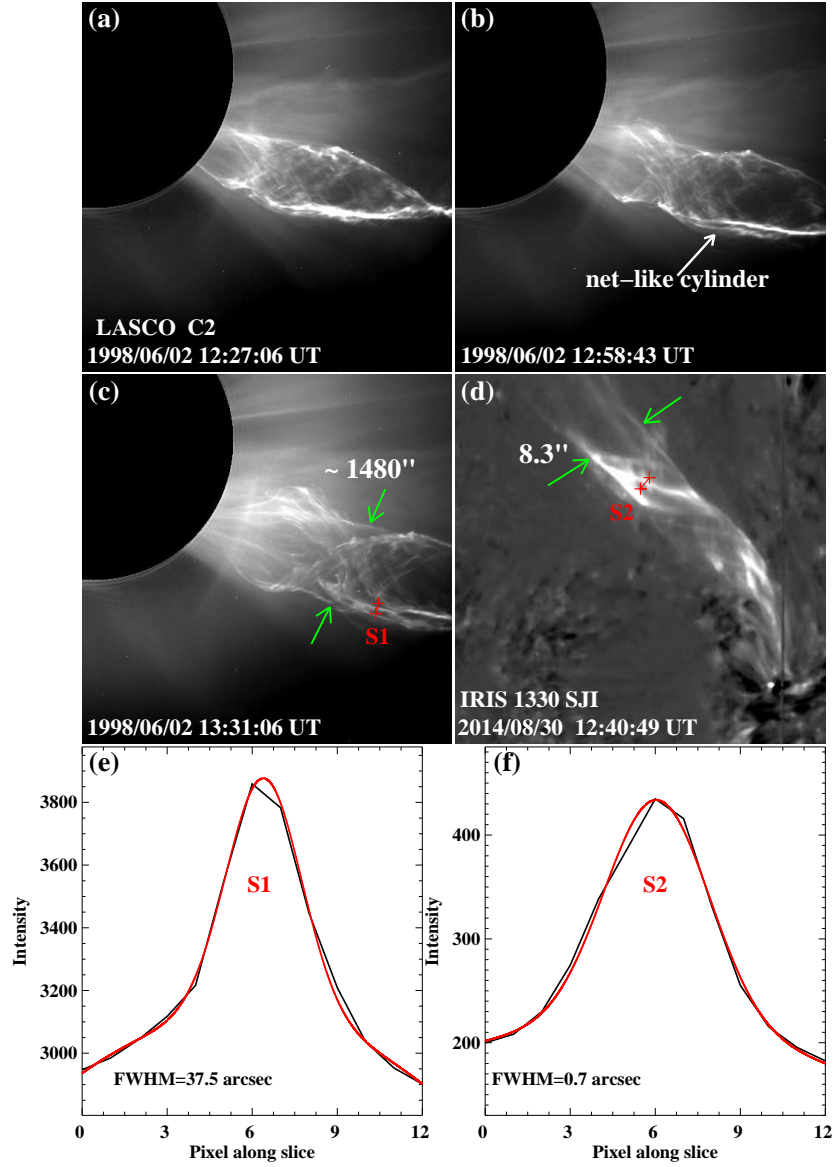}
              }
   \caption{LASCO/C2 images and IRIS 1330 SJI comparing
the CME core on 2 June 1998 with the flux rope, and the Gaussian
fitting profiles (panels (e) and (f)) showing the widths of
fine-scale structures. Green arrows in panels (c) and (d) point to
the location where the widths of the CME core and flux rope are
obtained. The red curves in panels (e) and (f) are respectively the
Gaussian fitting to the intensity profiles (black ones) along the
red slices (Slices ``S1" and ``S2") in panels (c) and (d).}
              \label{Fig7}%
    \end{figure}

At 12:37:03 UT, the line profile of location ``A" showed a redshift
of about 6 km s$^{-1}$ (panel (a)). Then the Doppler velocity
increased to 24 km s$^{-1}$ about 3 min later. At 12:40:11 UT, the
redshifts at locations ``B"$-$``D" were also observed with Doppler
velocity of 6$-$12 km s$^{-1}$ (panels (b)-(d)), smaller than that
of ``A". Moreover, the line profile of location ``B" at 12:37:03 UT
showed a blue-wing excess (panel (b)), and the minor component
contributing about 20\% to the total emission. The wing excess
corresponded to a strong blueshift of up to 44 km s$^{-1}$ (Peter,
2010). By investigating the SJIs, we found a jet-like activity
occurred between 12:35:10 UT and 12:37:59 UT. The plasma from the
jet possibly accounted for the blueshift at location ``B". For the
line profile of location ``C" at 12:37:03 UT, the emission intensity
rapidly increased to 200\% of the peak intensity at 12:35:29 UT
(blue line in panel (c)). Meanwhile, the line width increased
evidently. This might be caused by the transient heating process of
the local plasma.

\subsection{Comparison of the Flux Rope with the CME on 2 June 1998}

The morphology of the flux rope reminded us of a CME on 2 June 1998.
In LASCO/C2 observations, the CME core was composed of many
fine-scale structures, which were crossed and tangled with each
other (Figure 7(a)). The CME core had a cylindrical shape with
helical field lines wrapping around it (Figures 7(b) and (c)),
similar to the shape of the flux rope (Figure 7(d)). However, the
scales of the CME core and the flux rope were greatly different. The
CME core had a width of 1480$^{\prime\prime}$ (green arrows in
Figure 7(c)), equal to 1.54 $R_{\odot}$. The width of the flux rope
was approximately 8.3$^{\prime\prime}$ (green arrows in Figure
7(d)). We also measured the widths of fine structures (slices ``S1"
for the CME core and ``S2" for the flux rope in Figures 7(c)-(d)) by
using Gaussian function to fit the intensity profiles. The full
width at half maximum (FWHW) of the Gaussian fitting profile was
thought to be the width of fine-scale structure. The fine structure
of the CME core had a width of 37.5$^{\prime\prime}$ (Figure 7(e)),
about 4.5 times of the width of the entire flux rope
(8.3$^{\prime\prime}$). The fine structure of the flux rope was only
0.7$^{\prime\prime}$ wide, nearly 4 pixels of IRIS images.

\section{Summary and Discussion}

We report the observations of a flux rope at TR temperatures made by
IRIS on 30 August 2014 nearby AR 12149. The flux rope was located at
the southern part of an erupting filament based on the SDO
observations. Initially magnetic flux cancellation and obvious EUV
brightenings were observed. Then the filament was activated and the
bright material moved southward and filled in the body of the flux
rope observed at 1330 {\AA}. The fine structures of the flux rope
were tangled with each other, and the total twist was estimated to
be almost 4$\pi$. Afterwards, the flux rope exhibited a
counterclockwise unwinding motion (when viewed top-down) around its
main axis. Spectral observations displayed that the redshifts from
several to more than 20 km s$^{-1}$ appeared at the southern leg of
the flux rope in the late stage. The flux rope showed a helical
structure, similar to the CME core of 2 June 1998. However, the
spatial scales of the flux rope and the CME core are greatly
different.

Magnetic flux cancellation constantly occurred in about 5 h before
the appearance of the flux rope. We suggest that the magnetic flux
cancellation initiated reconnection and some activation of the flux
rope. The obvious EUV brightenings indicate that heating took place
during the filament eruption. The heated plasma from the filament
moved towards its southern end and illuminated the flux rope body in
the 1330 {\AA} channel. Thus the twisted structures of the flux rope
were tracked out by filling it with hot and dense plasma. We suggest
that part of the magnetic flux rope structures may exist in the
space filled in by the flux ropes before the appearance of the flux
rope. This is similar to the observations of Raouafi (2009) and Li
and Zhang (2013b), who reported the brightening of flux ropes and
the appearance of the fine structures due to the nearby flares or
other activities.

The filament material showed the counterclockwise rotation motion
along the helical flux rope structures in the late stage. The total
twist (4$\pi$) of the flux rope is above the critical value
(2.5$\pi$$-$3.5$\pi$; Hood and Priest, 1981), which indicates that
kink instability possibly took place in this event. The rotation was
obviously the unwinding motion of the twisted magnetic field lines.
The stored twist and magnetic helicity of the flux rope were
possibly transported to the ambient open fields (Liu \emph{et al.},
2009; Chen \emph{et al.}, 2012; Zhang \emph{et al.}, 2015). The
average angular speed (15.7$\times$10$^{-3}$ rad s$^{-1}$) of the
flux rope is comparable to that of the unwinding jet
(11.1$\times$10$^{-3}$ rad s$^{-1}$) presented by Shen \emph{et al.}
(2011). Koleva \emph{et al.} (2012) reported the twist of an
eruptive filament by 6$\pi$ and suggested that kink instability
played a key role in the filament eruption.

The IRIS slit covered the fine-scale structures at the western side
of the axis and the line profiles of C {\sc ii} at these locations
were mostly redshifted. The spectroscopic results are consistent
with the above inferred counter-clockwise rotations, with dominant
redshifts on the west. The redshifts at the northern locations are
larger than the south (in positions ``A" and ``D"). There might be
two reasons to explain the difference in the redshifts. The southern
locations were near to the footpoints of the flux rope and had a
smaller rotating radius. Moreover, the Doppler shifts at different
locations along the same helical structure are different due to the
projection effect, \emph{i.e.}, the points at the edge of the
helical structure had the largest Doppler shifts and the points in
the middle had the lower velocities. Su \emph{et al.} (2014)
revealed the opposite Doppler velocities of $\approx$ 5 km s$^{-1}$
at the two sides of the prominence, indicating the rotational motion
of the magnetic structures in tornado-like prominences. Liu \emph{et
al.} (2015) investigated an eruptive prominence by using IRIS
observations and found the blue to redshift transition with height
and sinusoidal spectral variations along the slit during the
counter-clockwise unwinding motions.

The small-scale flux rope (width of 8.3$^{\prime\prime}$) and the
large-scale CME core (width of 1.54 $R_{\odot}$) on 2 June 1998 both
showed a helical structure. They look very similar possibly because
of low optical density, and the fine-scale structures at both sides
of the flux rope are visible. This similarity shows the presence of
flux ropes of different scales on the Sun. The fine structure of the
flux rope is about 0.7$^{\prime\prime}$ wide, similar to the
previous results. Li and Zhang (2013b) showed that the flux ropes in
EUV wavelengths are composed of about 100 fine-scale structures,
with an average width of about 1.6$^{\prime\prime}$. Yang \emph{et
al.} (2014) analyzed a flux rope in H$\alpha$ images and measured
the average width of its individual threads of 1.11 Mm. The fine
structure of the CME core has a width of about
37.5$^{\prime\prime}$, much thicker than that of the flux rope.

The comparison of multi-wavelength observations showed that some
threads of the flux rope emitted only in spectral lines at TR
temperatures, \emph{e.g.}, C {\sc ii}. The structures of the flux
rope could not be clearly observed in high-temperature wavelengths
such as 171 and 304 {\AA}. This suggests that some plasma of the
flux rope is heated to only 25000 K, rather than a higher
temperature. The comprehensive characteristics of flux ropes at TR
temperatures need to be analyzed in further studies, together with
the comparison of low- and high-temperature flux ropes.

\begin{acks}
IRIS is a NASA small explorer mission developed and operated by
LMSAL with mission operations executed at NASA Ames Research center
and major contributions to downlink communications funded by the
Norwegian Space Center (NSC, Norway) through an ESA PRODEX contract.
We acknowledge the SDO/AIA and HMI for providing data. This work is
supported by the National Basic Research Program of China under
grant 2011CB811403, the National Natural Science Foundations of
China (11303050, 11025315, 11221063 and 11003026), the CAS Project
KJCX2-EW-T07 and the Strategic Priority Research Program$-$The
Emergence of Cosmological Structures of the Chinese Academy of
Sciences, Grant No. XDB09000000.
\end{acks}

{}

\end{article}

\end{document}